\documentstyle[prl,aps,psfig]{revtex}

\newcommand{\be}{\begin{equation}}
\newcommand{\ee}{\end{equation}}
\newcommand{\eq}[1]{Eq.~(\ref{#1})}

\newcommand{\fig}[1]{Fig.~\ref{#1}}

\begin{document}
\preprint{\bf PREPRINT}
\columnsep0.1truecm
\draft

\twocolumn[\hsize\textwidth\columnwidth\hsize\csname@twocolumnfalse\endcsname

\title{Island Distance in One--Dimensional Epitaxial Growth}
\author{Harald Kallabis$^{1,2}$, Paul L. Krapivsky$^2$ 
and Dietrich E.~Wolf$^{1,2}$}
\address{
$^1$ FB 10, Theoretische Physik, Gerhard--Mercator--Universit\"at Duisburg,
47048 Duisburg, Germany\\
$^2$ Center for Polymer Studies, Boston University, Boston, 02215 MA, USA
}

\maketitle

\begin{center}
\today
\end{center}

\begin{abstract}
The typical island distance $\ell$ in submonlayer epitaxial growth
depends on the growth conditions via an exponent $\gamma$. This
exponent is known to depend on the substrate
dimensionality, the dimension of the islands, and
the size $i^*$ of the critical nucleus for island formation.  In this paper
we study the dependence of $\gamma$ on $i^*$
in one--dimensional epitaxial growth.  We derive
that $\gamma = i^*/(2i^* + 3)$ for $i^*\geq 2$ and confirm this result by 
computer simulations.

\end{abstract}

]

\narrowtext

\section{INTRODUCTION}

Molecular beam epitaxy (MBE) is of outstanding importance for the
fabrication of microelectronic devices. The highly controllable
experimental conditions make this technique an ideal tool for crystal
growth. The thickness of the deposited layers can be controlled down
to the atomic scale while growth proceeds layerwise.

Typically the growth of multiple layers is of primary interest in most
experimental and theoretical studies of MBE. However, deeper insight into
the physics of multilayer growth requires a comprehensive understanding of 
the submonolayer regime (see
Ref.~\cite{Damping}, for instance).  The most important length scale in
submonolayer growth is the island distance or diffusion length $\ell$.
The classical result \cite{Gamma} for this length is that it depends
on the diffusion constant $D$ of adatoms and the deposition rate
$F$ like

\be
\ell \sim (D/F)^\gamma.
\label{ellD}
\ee
The analysis of rate equations yields that the exponent $\gamma$
depends on the substrate dimension $d$, on the size $i^*$ of the
critical nucleus and the fractal dimension $d_f$ of the
islands \cite{Gamma,WolfNATO,Seoul}:

\be
\label{gamma}
\gamma = \frac{i^*}{2i^*+d+d_f}.
\ee
The critical nucleus is defined such that all islands consisting of
$i^*$ or less atoms can shrink by emitting adatoms, while an island of
size $i^*+1$ is stable. Eq. (\ref{gamma}) applies if island diffusion
and adatom desorption are not important.

\eq{gamma} has been demonstrated to hold in $d=2$ for
$i^*=1,2,3,4$, see Refs.~\cite{Schroeder,Jeong}. In this paper we
study the dependence (\ref{ellD}) in surface dimension $d=1$ 
for different values of the
critical nucleus $i^*$. In this case, $d_f=1$. So far, only the case
$i^*=1$ \cite{WolfNATO,PVW} has been studied in one dimension, to our
knowledge, and it is in agreement with (\ref{gamma}). Here we show that
(\ref{gamma}) is no longer valid for $i^* \geq 2$, where 

\be
\label{gamma1}
\gamma = \frac{i^*}{2i^*+3}.
\ee

Our motivation for studying the one-dimensional case is the following.  
Much theoretical work is dedicated to the study of basic
mechanisms of epitaxial growth. For computational reasons,
the numerical work in this area is often restricted to models in
surface dimension $d=1$. As mentioned above, such studies rely on a
good understanding of the physics in the submonolayer regime. 

Also, quasi one--dimensional MBE may be realized experimentally on
surfaces with extremely anisotropic diffusion constant, like the
2x1--dimer--reconstructed Si(001) surface (see, e.g.,
Ref.~\cite{Lagally}).  There, the estimated ratio of the diffusion
constants along rows and perpendicular to them is of the order of
1000:1 in typical experiments \cite{Mo}. A computer simulation study
of this growth process leads to the estimate that for a deposition
rate of $F=1/600$~ML/s and at temperatures below $T=450$~K, this
anisotropy should be pronounced enough to lead to quasi
one--dimensional MBE \cite{SiEvans}. However, there is a controversy
concerning the relative importance of anisotropic diffusion and
anisotropic bonding \cite{Clarke,Pearson}. Another experimental
example is the reconstructed surface of Au(100) \cite{au}.

This paper is organized as follows.  In section II the
expression (\ref{gamma1}) will be derived. The subsequent sections are devoted
to the numerical confirmation of this result. In section III we employ the
coarse--grained MBE (CGMBE) model \cite{WolfNATO}, which allows for a simple
algorithmic implementation of a variable critical nucleus size $i^*$. To
ensure that the results are independent of the chosen algorithm, we 
reproduce them
with a model that is not coarse grained (MBE model) in
section IV.  It is shown that the results of the MBE and the CGMBE models
compare favourably with each other and with the theoretical prediction.  The
final section V contains a general discussion of the results.

\section{ANALYTICAL CONSIDERATIONS}
\label{analaytical.sect}

Most treatments of the monolayer growth processes are based on a mean-field
rate equation approach. This is appropriate in the most important
two--dimensional case since the basic kinetic mechanism -- the two-particle
diffusion-controlled aggregation -- has an upper critical dimension
$d_c=2$\cite{van}.  In one dimension, however, classic rate equations do not
apply and we shall employ a modified rate equation approach \cite{PVW}.

We denote by $A$ and $I$ densities of adatoms and stable immobile
islands, respectively. We start with adatoms and write
\begin{equation}
\label{mon}
{dA\over dt}=F-{A\over \tau}.
\end{equation}
Here $\tau$ is a lifetime, {\it i.e.}, the time between the deposition of
an adatom and its incorporation into a stable island.
During the time interval $\tau$, the adatom
visits $\sqrt{D\tau }$ different sites, so the adatom lifetime is determined
by $I\sqrt{D\tau }\approx 1$, and Eq.~(\ref{mon}) becomes
\begin{equation}
\label{mono}
{dA\over dt}=F-DI^2A.
\end{equation}
This differs from the two--dimensional situation where the number of
different sites visited by an adatom grows linearly with time and the adatom
density evolves according to $dA/dt=F-DIA$. Actually, Eq.~(\ref{mono}) contains
an additional loss term due to the nucleation of new islands. 
However, this loss term is asymptotically subdominant, since most 
adatoms are captured by the existing islands. 

The rate equation for the island density is derived by noting that
islands are formed upon colliding of $i^*+1$ adatoms. Hence
\begin{equation}
\label{island}
{dI\over dt}=DA^{i^*+1}.
\end{equation}
The many-particle diffusion-controlled reaction 
process $(i^*+1)A\to I$ has an upper
critical dimension $d_c=2/i^*$ (see e.g.\cite{pk} and references therein).
The fact that the upper critical dimension gets smaller when $i^*$ increases
is simple to appreciate. Indeed, many-particle collisions occur rarely, so
the system is well mixed and the mean-field approach should become correct
earlier. For $i^*>2$ we have $d_c<1$, so Eq.~(\ref{island}) applies; the case
$i^*=2$ is marginal, so we anticipate logarithmic corrections to the
mean-field results. For $i^*=1$ correlations dominate the diffusion process
in one dimension, so that (\ref{island}) has to be modified \cite{van} and
leads to $\gamma=1/4$ \cite{PVW}.

Eq.(\ref{gamma1}) is obtained from (\ref{island}) and (\ref{mono}) in the 
following way: From (\ref{mono}) it follows that asymptotically
$A\simeq (F/D)I^{-2}$. Inserting this into (\ref{island}) and intergrating
we find 
\begin{equation}
\label{it}
I\sim \left({F\over D}\right)^{i^*/(2i^*+3)}(Ft)^{1/(2i^*+3)}.
\end{equation}
Setting $t\propto F^{-1}$ gives the result (\ref{gamma1}).

For $i^*=2$ there is a logarithmic correction to (\ref{island}) \cite{pk}, 
\begin{equation}
\label{island1}
{dI\over dt}={DA^{3}\over |\ln I|},
\end{equation}
whereas (\ref{mono}) remains unchanged. Repeating the same analysis gives
\begin{equation}
\ell\sim \left( {D\over F}\right)^{2/7} 
\left[\ln{\left({D\over F}\right)}\right]^{1/7}.
\end{equation}

\section{NUMERICAL RESULTS}
\label{CGMBE.sect}

The idea of the coarse grained MBE (CGMBE) model, introduced in
Ref.~\cite{WolfNATO}, successfully applied in Refs.~\cite{Schroeder,Jeong},
and studied in detail in Ref.~\cite{KallDiss}, is to avoid the time-consuming
simulation of the surface diffusion on the atomic scale by resolving it only
on a much coarser scale $\Delta x$. As long as $\Delta x$ is smaller than the
island distance $\ell$, one still gets correct information about the surface
morphology.

Specificially the model is implemented as follows: The substrate is
divided into cells of linear extension $\Delta x$, so that a monolayer
corresponds to $\Delta x$ atoms per cell. The partial filling of a
cell at site $x$ is given by a counter $m(x)=0,1,\dots,\Delta x-1$,
which means the number of atoms on top of completed monolayers in cell
$x$. The height $h(x)$ denotes the number of complete monolayers. For
the complete description of the state of a cell, an edge indicator
$is(x)$ is introduced. If $is(x)=1$, atoms cannot leave cell $x$
because they are irreversibly incorporated into the island edge
present in the cell. This allows for a simple implementation of the
concept of the critical nucleus $i^*$: if upon diffusion or deposition
of an atom into cell $x$ the counter reaches the value $m(x)=i^*$,
$is(x)$ is set to one. Consequently, $m(x)$ can only increase from
then on, until $m(x)=\Delta x$, in which case $h(x)$ is increased by
one and $m(x)$ and $is(x)$ are set to zero. As long as $is(x)=0$,
$m(x)$ can also decrease, when atoms diffuse into neighbouring cells.

The dynamics has to take into account that a single adatom move from
one cell to a neighbouring one represents the diffusion over a
distance of $\Delta x$ lattice sites.  This happens with frequency
$\nu_{\rm D}=D/(\Delta x)^2$ for all adatoms in the system. The
deposition of one atom happens with frequency $\nu_{\rm F}= F L \Delta
x$. Hence the probability for deposition during one time step $\Delta
t = (\nu_{\rm F} + \nu_{\rm D})^{-1}$ of the computer simulation is

\be
\label{CGMp}
p=\nu_{\rm F}/(\nu_{\rm F} + \nu_{\rm D}).
\ee
$1-p$ is the probability with which each of the adatoms is allowed to
diffuse into a neighboring cell. If no adatoms are present, $\nu_{\rm D}=0$.
This algorithm speeds the simulation up by a factor of about $(\Delta
x)^2$.

In the simulations presented in the following, the coarse graining
length had the constant value $\Delta x=20$. The system size was
chosen to be $L\Delta x=2\cdot 10^5$ and averages were taken over 100
independent runs.

\begin{figure}
\centerline{\psfig{figure=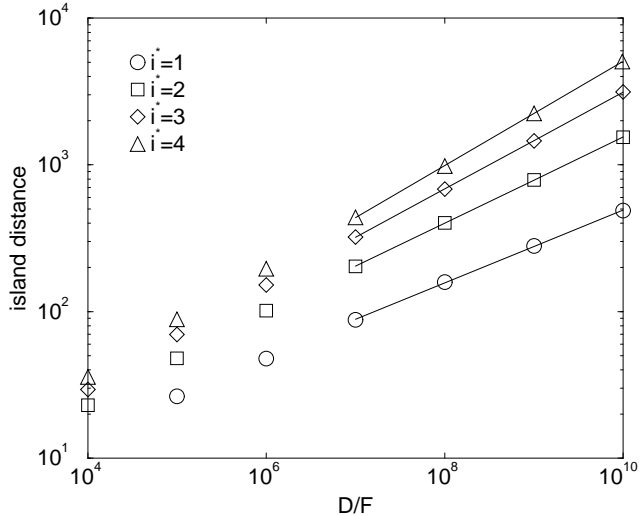,height=7.0cm,angle=270}}
\caption{
  The island distance as a function of $D/F$ for $i^*=1,2,3,4$.  The
  solid lines represent fits to the last three decades of data and
  have the slopes $\gamma=0.248, 0.293, 0.329, 0.355$, respectively.
}
\label{lCGMBE.fig}
\end{figure}

In the CGMBE model, nucleation events can be identified easily.  Thus
the island distance can be determined by dividing the system size by
the total number of nucleation events in the first layer. Measurements
of this quantity as a function of $D/F$ for different $i^*$ are shown
in \fig{lCGMBE.fig}.  Note that the asymptotic algebraic behaviour
$\ell\sim(D/F)^\gamma$ is reached with $\ell$ as function of $D/F$
having a weak negative curvature.

\section{RESULTS WITHOUT COARSE GRAINING}
\label{nocg.sect}

In the (non--coarse grained) MBE model atoms are deposited onto the
surface of a simple cubic crystal (i.e.\ a square lattice in $d=1$)
with a rate of $F$ atoms per unit time and area.  Atoms with no
lateral neighbors are allowed to diffuse with diffusion constant
$D$. In the simplest variant of the model atoms with lateral neighbors
are assumed to be immobile so that, e.g., dimers are immobile and
stable. Therefore, $i^*=1$.  Extension of the rules from $i^*=1$ to
$i^*=2$ is straightforward: Dimers are allowed to split into two
adatoms, while trimers are stable.  In either case, growth commences
with a flat substrate, $h(x,0)=0$ for all sites $x$.  On deposition at
$x$, $h(x,t)$ is increased by one.  We neglect barriers to interlayer
transport (Ehrlich--Schwoebel barriers \cite{ESB}).  Thus the only
parameter of the model is the ratio $D/F$.  This model was simulated
in systems of size $L=65536$.

The island distance was measured here as follows. During deposition,
the number of islands was calculated for times $0<Ft<1$. An island was
defined as a connected region where $h(x)\ge a_\perp$, where $a_\perp$
is the vertical lattice constant.  For early times, the number of
islands increases due to nucleation. At roughly $Ft\simeq0.5$ the
number of islands decreases again since the islands coalesce. This
means that the island distance has a minimum which was used for
evaluation.

\begin{figure}[htb]
\centerline{\psfig{figure=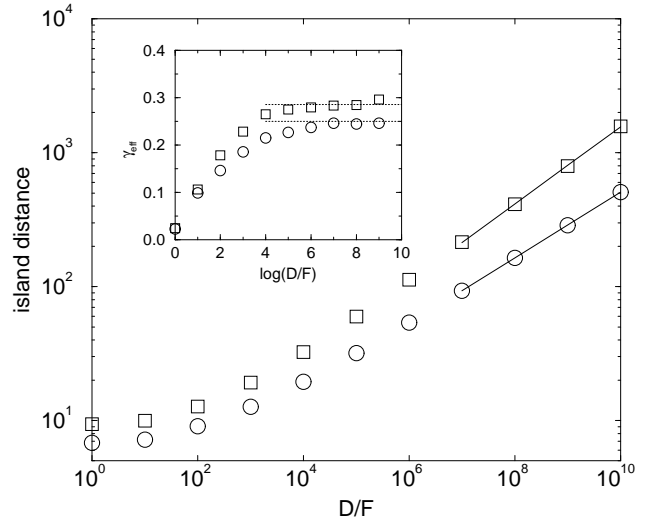,height=7.0cm,angle=270}}
\caption{
  The island distance as a function of $D/F$ for $i^*=1$ (circles) and
  $i^*=2$ (squares). The solid lines with slopes $\gamma=0.246$ and
  $\gamma=0.288$ represent fits to the last three decades of data.
  The inset shows how the local slopes approach the theoretically 
  predicted exponents (dotted lines) for large $D/F$.
}
\label{l.fig}
\end{figure}

In \fig{l.fig}, the measurements of the island distance as a function
of $D/F$ for the MBE model with $i^*=1$ and $i^*=2$ are shown. From
the effective exponent $\gamma_{\rm eff}\equiv \log(10\cdot
D/F)-\log(D/F)$ shown in the inset, it becomes clear that the
asymptotic behaviour $\ell\sim(D/F)^\gamma$ is reached only for 
large values of $D/F$.  The dotted lines indicate the theoretical
values from \eq{gamma}, $\gamma(i^*=1)=1/4$, and \eq{gamma1}, 
$\gamma(i^*=2)=2/7$,
respectively.  Note also that the asymptotic exponent is approached from
below.

\section{DISCUSSION}
\label{discussion.sect}

\begin{figure}[htb]
\centerline{\psfig{figure=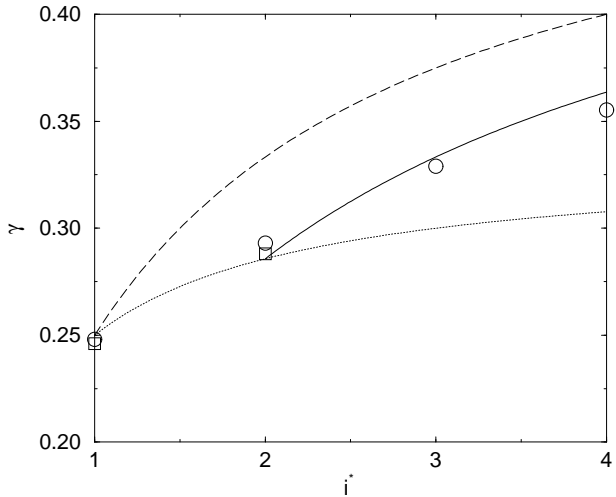,height=7.0cm,angle=270}}
\caption{
  The submonolayer exponent $\gamma$ as a function of $i^*$ in
  $d=1$. Squares
  and diamonds are the  results of the simulations of the
  CGMBE and the MBE models, respectively.  
  The solid curve is the theoretical prediction for $i^* \geq 2$, 
  \protect\eq{gamma1}.
  The dashed curve corresponds to formula (\protect\ref{gamma}) and the 
  dotted line are the exponents expected, if islands up to size $i^*$ 
  can diffuse, \protect\eq{gamma2}.  
}
\label{gamma.fig}
\end{figure}

In \fig{gamma.fig}, the comparison of the simulation results of the
CGMBE model (\fig{lCGMBE.fig}) and the MBE model (\fig{l.fig}) with
various analytical formulas is shown.  
The small difference between
$\gamma$ measured in the CGMBE and the MBE model is not only due to
statistical noise, but also due to the asymptotic behaviour of
$\ell(D/F)$ being reached with negative and positive curvature,
respectively. Therefore, even better agreement between the results
of the CGMBE and the MBE models for larger $D/F$ can be expected.  As
a consequence, the values for higher $i^*$, obtained with the CGMBE
model, can be regarded as reliable.

The agreement of the numerical results with the theoretical prediction
(\ref{gamma1}) is very good. For comparison, we also show Eq. 
(\ref{gamma}), which in one dimension applies only for $i^*=1$. 
It is interesting to 
notice that the exponent for $i^*=2$ coincides with the one expected,
if islands up to size $i^*$ can diffuse without decaying
\cite{WolfNATO,Biham}
 
\be
\gamma = \frac{i^*}{(d+2)i^*+d_f} = \frac{i^*}{3 i^* + 1}.
\label{gamma2}
\ee
In fact, unstable islands effectively diffuse in our models.
Consider for example $i^*=2$, where dimers are
unstable. When a dimer decays, the adatoms still remain close to each
other, and in one dimension the probability is particularly high that
they reunite again at a shifted position.  The specific way we
implemented the decay of a cluster in our computer simulations makes
this tendency very clear: Whenever a dimer has formed it must break up
in the next diffusion step, which means that each of the two atoms
moves to one of the neighboring cells. With probability $1/2$ they
will be found in the same cell. If no further adatom was met there,
this amounts simply to a displacement of the dimer by $\pm \Delta x$.
This analogy breaks down, however, for $i^*>2$.

In conclusion, we have investigated the submonolayer epitaxial growth in a
one dimensional model where islands of size up to $i^*$ are unstable while larger islands are
stable and immobile.  We have derived the exponent $\gamma$ which determines
the $D/F$--dependence of the diffusion length in one dimension for critical
island size $i^*\geq 2$ and have confirmed the result by computer
simulations.

\section{Acknowledgements}

D.~E.~W.~ acknowledges support by DFG within SFB 166. 
H.~K.~ acknowledges support by DAAD within the
Hochschulsonderprogramm III.
P. L. K. acknowledges NSF grant DMR-9632059 and ARO grant
DAAH04-96-1-0114 for financial support.

\section{Note added in proof} 
Following the suggestion of one of the referees we note that our rate
equation approach applies to the general case of surface dimension $d$
and island dimension $d_f$.  The fractal island dimension can be
smaller or larger than the surface dimension which is integer in
physically relevant cases.  In this general case we use $F t \ell^d =
\ell^{d_f}$ with $\ell=I^{-1/d}$ to obtain

\[
\gamma = \frac{i^*}{d\nu(i^* + 1) + d_f},
\]
where $\nu=2$ for $d=1$ and $\nu=1$ for $d\ge2$. This does not apply
to the exceptional case $d=i^*=1$ where $d<d_c=2/i^*$.  There equation
(6) should read $\dot I = DA^2I$ and one finds $\gamma = 1/(3 + d_f)$.

\end{document}